\documentclass[usenatbib,usegraphicx]{mn2e}

\usepackage{diagbox} % for diagonal line in tables
\usepackage{amsmath}
\usepackage{graphicx}
\usepackage{subcaption}
\usepackage[
pdfauthor={Barbara Ercolano},
pdftitle={Photoevaporation for realistic X-ray spectra},
pdfstartview=FitH,
linkcolor=blue,
anchorcolor=blue,
citecolor=blue,
filecolor=blue,
menucolor=blue,
urlcolor=blue,
colorlinks=true]{hyperref}
\usepackage{longtable}
\usepackage{aas_macros}
\usepackage{siunitx}
\usepackage{amssymb}
\usepackage{graphicx}  %for images
\graphicspath{{plots/}} 

\usepackage{caption}
\DeclareCaptionLabelFormat{cont}{Figure#1~#2\alph{ContinuedFloat}}
\usepackage{xcolor}

\DeclareSIUnit\Msol{M_\odot}   % Solar mass
\DeclareSIUnit\Zsol{Z_\odot}   % Solar metallicity
\DeclareSIUnit\Gnot{G_0}       % Draine radiation field
\DeclareSIUnit\Myr{Myr}        % Million years
\DeclareSIUnit\kyr{kyr}        % thousand years
\DeclareSIUnit\mag{mag}        % magnitude
\DeclareSIUnit\au{au}          % au

\newcommand\ergs{\mathrm{erg\,s^{-1}}}
\newcommand\Lx{L_\mathrm{X}}
\newcommand\Mdotwind{\dot{M}_\mathrm{w}}
\newcommand\Sigmadotwind{\dot{\Sigma}_\mathrm{w}}
\newcommand\gcm{\mathrm{g\,cm^{-2}}}
\newcommand\au{\mathrm{au}}
\newcommand\Msun{M_\odot}

\newcommand\spock{\textsc{spock}}
\newcommand\mocassin{\textsc{mocassin}}
\newcommand\pluto{\textsc{pluto}}

\defcitealias{Picogna2019}{Paper~I}

\title[]{The dispersal of protoplanetary discs. -- II: Photoevaporation models with observationally derived irradiating spectra}
\author[Ercolano et al.]{{Barbara Ercolano}$^{1,2}$\thanks{E-mail: ercolano@usm.lmu.de},
       {Giovanni Picogna}$^{1}$, 
       {Kristina Monsch}$^{1,3}$, 
       {Jeremy J. Drake}$^{3}$, \and
       {Thomas Preibisch}$^{1}$\\
       $^{1}$ Universit\"ats-Sternwarte, Fakult\"at f\"ur Physik,   Ludwig-Maximilians-Universit\"at M\"unchen, Scheinerstr.~1, 81679 M\"unchen, Germany\\
       $^{2}$ Exzellenzcluster `Origins', Boltzmannstr.~2, 85748 Garching, Germany \\
       $^{3}$ Smithsonian Astrophysical Observatory, 60 Garden Street, Cambridge MA02138, USA\\}

\newenvironment{itemize*}%
  {\begin{itemize}%
    \setlength{\itemsep}{0pt}%
    \setlength{\parskip}{0pt}}%
  {\end{itemize}}

\begin{document}

\maketitle

\begin{abstract}

Young solar-type stars are known to be strong X-ray emitters and their X-ray spectra have been widely studied. X-rays from the central star may play a crucial role in the thermodynamics and chemistry of the circumstellar material as well as in the atmospheric evolution of young planets. In this paper we present model spectra based on spectral parameters derived from the observations of young stars in the Orion Nebula Cluster from the \textit{Chandra} Orion Ultradeep Project (COUP). 
The spectra are then used to calculate new photoevaporation prescriptions that can be used in disc and planet population synthesis models. Our models clearly show that disc wind mass loss rates are controlled by the stellar luminosity in the soft ($100\,\mathrm{eV}$--$1\,\mathrm{keV}$) X-ray band. New analytical relations are provided for the mass loss rates and profiles of photoevaporative winds as a function of the luminosity in the soft X-ray band. 
The agreement between observed and predicted transition disc statistics moderately improved using the new spectra, but the observed population of strongly accreting large cavity discs can still not be reproduced by these models. Furthermore, our models predict a population of non-accreting transition discs that are not observed. This highlights the importance of considering the depletion of millimeter-sized dust grains from the outer disc, which is a likely reason why such discs have not been detected yet. 
\end{abstract}


% ----------------------------------------
% ----------------------------------------
% ----------------------------------------
\section{Introduction} 
\label{sec:intro}

Thermal and magnetic winds are thought to play a crucial role in the evolution of protoplanetary discs \citep[e.g.][]{2017RSOS....470114E, Alexander+2014}. In particular, observations suggest that the final dispersal of the disc material must occur quickly and proceed from the inside-out \citep[e.g.][]{Koepferlf+2013, Ercolano+2015, Luhman+2010, ErcolanoClarkeHall2011}. Thermally unbound and centrifugally launched winds, known as photoevaporative winds, can easily reproduce this observed fast, inside-out dispersal, and in particular, X-ray-driven photoeavaporative winds have been shown to produce mass loss rates comparable to observed accretion rates on young low-mass stars \citep{Ercolano+2009, Owen_2010, Picogna2019}. Also, a significant fraction of the so-called transition discs can be explained either by disc dispersal via photoevaporation alone \citep{ErcolanoWeber+2018} or in combination with magnetic fields \citep{WangGoodman2017a}. 

There is growing indirect evidence that X-rays from the central star may indeed be the major driver of photoevaporative winds \citep[e.g.][]{Ercolano+2014, Monsch+2019, Flaischlen+2021}, and these models can well reproduce some of the observed emission line diagnostics \citep{ErcolanoOwen2010, ErcolanoOwen2016, Weber+2020}, although recent observations point to the presence of additional emission components which might also be attributed to magnetic disc winds \citep[e.g.][]{Pascucci+2020, Banzatti+2019, Nisini+2018, Gangi+2020}.

Current X-ray photoevaporation models span the observed parameter space in X-ray luminosities \citep[][hereafter \citetalias{Picogna2019}]{Picogna2019}, carbon depletion \citep{Woelfer_2019}, and stellar masses (Picogna et al., in preparation). All of these models, however, use the same irradiating spectrum, namely the synthetic spectrum employed by \citet{Ercolano+2008b, Ercolano+2009}. Other studies, that employed different irradiating spectra in the X-ray domain have therefore yielded different conclusions as to the efficacy of X-rays to drive a photoevaporative wind. In particular, when using hard X-ray spectra with substantial flux at or above $1\,\mathrm{keV}$, other authors have found that X-rays become inefficient at driving the wind \citep{WangGoodman2017b, Nakatani+2018}. 
This is not surprising as was shown by \citet{Ercolano2009}, who found that the soft X-ray band ($< 1\,\mathrm{keV}$) by far dominates the heating in the wind launching regions. While hard X-rays can penetrate larger columns of gas where the densities are much higher, they are nevertheless unable to provide enough heating in those regions to unbind the gas.
Thus it is clear that a realistic X-ray spectrum is of crucial importance to determine the thermodynamics in the disc atmospheres. This affects the wind as well as the emission line spectrum emerging from the disc atmosphere and the wind \citep{Schisano+2010}. 

In this paper we use the results from the \textit{Chandra} Orion Ultradeep Survey \citep[COUP, cf.][]{Getman+2005} of young stars in the Orion nebula cluster to infer typical parameters of the X-ray spectrum. Synthetic spectra are then derived from these parameters and used as input for radiation-hydrodynamical simulations of photoevaporating discs. We derive mass loss rates and profiles and provide prescriptions for population synthesis codes. The effects of the new calculations on the evolution of the disc surface density as well as the formation of transition discs and the migration of planets is also explored. 

The new input spectra are derived and described in Section~\ref{sec:Xray}, while in Section~\ref{sec:methods} we summarise the numerical methods employed in the paper. Our results are presented in Section~\ref{sec:results}. Finally, Section~\ref{sec:conclusion} contains a discussion and a brief summary of the conclusions of this work.

% ----------------------------------------
% ----------------------------------------
% ----------------------------------------
\section{The X-ray spectra of T Tauri stars} 
\label{sec:Xray}

\begin{table}
\noindent
\begin{tabular}{ccccc}
\hline
$\Lx$  & $T_1$ & $T_2$ & $EM_2/EM_1$ & model name \\  
(erg/s) & ($10^6\,$K) & ($10^6\,$K) & & \\
\hline
$10^{29}$    &     8  &      22    &   0.6 & \texttt{Spec29}\\

$10^{30}$    &     9  &      30    &   1.6 & \texttt{Spec30}\\

$10^{31}$    &    10  &      35    &   2.5 & \texttt{Spec31}\\
\hline
\end{tabular}
\caption{Mean spectral parameters from two-component temperature fits of some 500 X-ray spectra of T~Tauri stars in the COUP data. $T_1$ and $T_2$ correspond to the temperatures of the ``cool'' and ``hot'' plasma components, respectively, while $EM_2/EM_1$ indicates the ratio of the corresponding emission measures. The spectral hardness increases with increasing X-ray luminosity.}
\label{tab:spec}
\end{table}

\begin{figure}
\centering
\includegraphics[width=\linewidth]{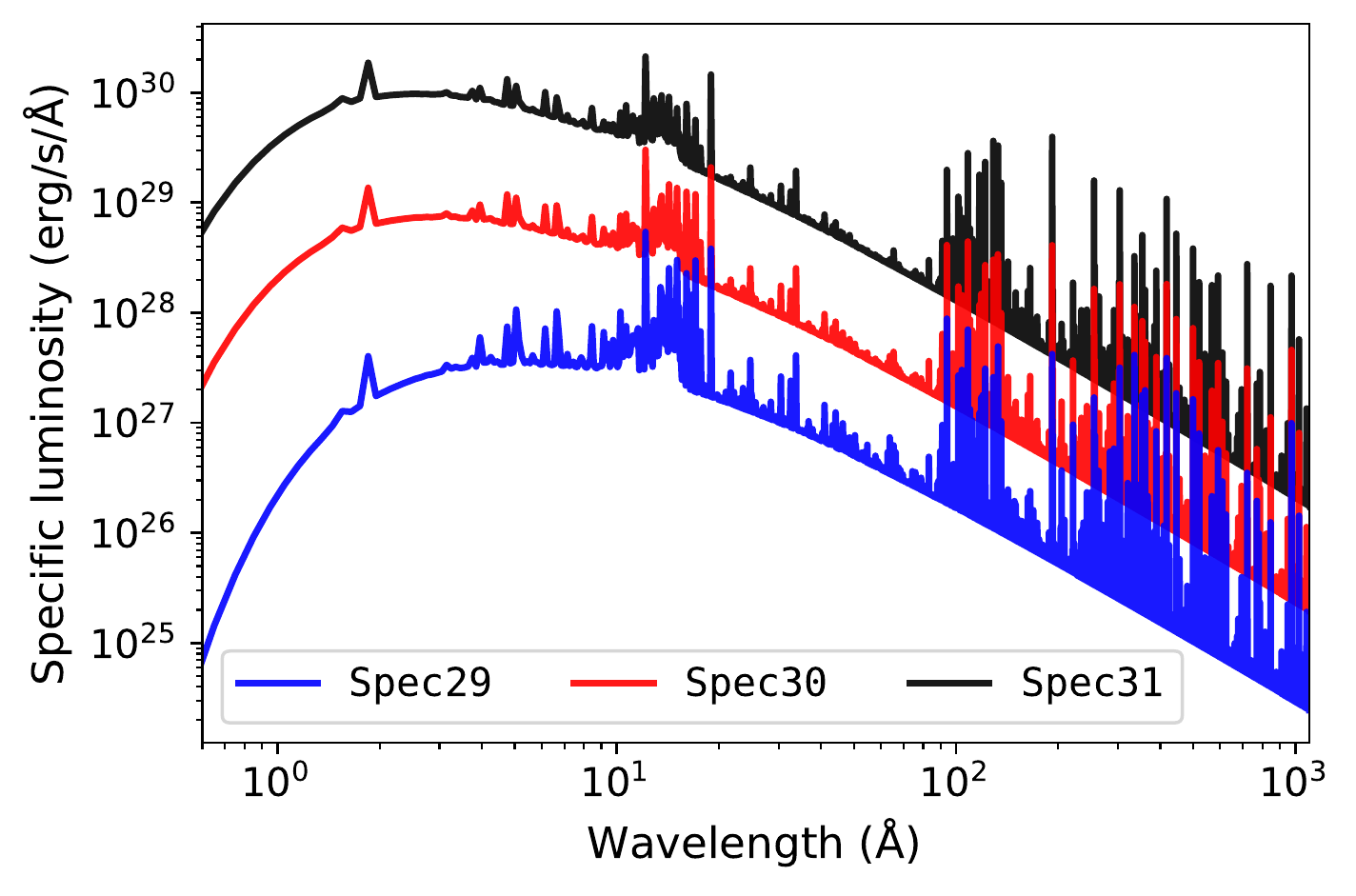}
  \caption{Synthetic spectra representative for the low-, medium- and high-X-ray luminous stars presented in this study, generated from observationally derived emission measures from the COUP data (cf. Section~\ref{sec:Xray} and Table~\ref{tab:spec} for details).}
     \label{fig:Xray}
\end{figure}

For the simulations in this study, we calculated X-ray spectra appropriate for T~Tauri stars with
different levels of X-ray activity. The spectral parameters for these models were based on the results of the COUP project,
an 838 ksec long \textit{Chandra} observation of the Orion Nebula Cluster (ONC) and the deepest X-ray observation ever made of a young stellar cluster.
The COUP data analysis showed that for the large majority of sources, the X-ray spectra could be well fitted by two-temperature thermal plasma models plus absorption. 
The resulting spectral parameters were thus:
($a$) the temperatures of the cool and hot plasma components, $T_1$ and $T_2$,
($b$) the emission measures $EM_1$ and $EM_2$ of both components, and
($c$) the hydrogen column density for the foreground extinction
\citep[see][for details of these parameters]{Getman+2005}.

\citet{Preibisch+2005} performed an analysis of the X-ray properties of the 598 optically visible likely ONC members that were detected as X-ray sources in the COUP data.
This showed that the temperature of the cool plasma component ($T_1$)
slightly increases with X-ray luminosity (or X-ray surface flux),
while the hot plasma component ($T_2$)
displays a stronger and clearer increase with X-ray luminosity (or X-ray surface flux).
It was also found that the ratio of the emission measures, $EM_2 / EM_1$,
systematically increases with the X-ray luminosity.

In order to cover the range of ``typical'' X-ray spectral properties
for stars with different values of X-ray luminosity, we determined the typical values for $T_1$, $T_2$, and the ratio $EM_2 / EM_1$ for three values of the X-ray luminosity, namely $\log \left( L_{\rm X}/\mathrm{erg\,s^{-1}} \right) =  29, \,
30,\, {\rm and}\, 31$ from the COUP data. These values are thus
approximately representative of low-, medium- and high-X-ray luminous stars
and thus cover the region in parameter space where the large majority of all
T~Tauri stars are located. This resulted in the values presented in Table~\ref{tab:spec}.

%The temperature of coronal plasma on late-type stars in general,
%including T~Tauri stars, shows a positive correlation with the stellar
%X-ray luminosity \citep{??refs??}. In order to simulate the EUV-X-ray
%spectra of disc-bearing stars, we first determined typical values of
%plasma temperature for three values of X-ray luminosity spanning the
%typical valuese exhibited by T~Tauri stars: $log(\Lx) = 29$, 30, and 31
%erg~s$^{-1}$ \citep[e.g.][]{??refs??}.  These $\Lx$ values are
%approximately representative of  low, medium and high X-ray luminosity
%stars.

%\citet{??ref??} have performed fits to {\it Chandra} X-ray spectra of
%several hundred T~Tauri stars in Orion based on the COUP survey
%\citep{Feigelson??} using two-temperature optically-thin,
%collision-dominated plasma radiative loss models. Taking these results
%and averaging them for our three adopted representative X-ray
%luminosities, we find the mean two-temperature fit results listed in
%Table~\ref{t:2temp}.

Finally, we computed the X-ray spectra based on the plasma temperatures and
relative emission measures shown in Figure~\ref{fig:Xray} using the {\sc
pint}of{\sc ale} IDL software package\footnote{available at \href{https://hea-www.harvard.edu/PINTofALE/}{https://hea-www.harvard.edu/PINTofALE/}.} \citep{KashyapDrake2000} using
spectral line and continuum emissivities computed from the {\sc
CHIANTI} database version 8.0.6. Spectra were then normalised to our adopted values of X-ray luminosity in the energy range of 0.5--5\,keV.

\section{Photoevaporation Calculation Methods} 
\label{sec:methods}

% ----------------------------------------
We follow the approach of \citet{Owen_2010,Owen_2011,Owen_2012} and \citetalias{Picogna2019}, and perform 2D radiation-hydrodynamical calculations of primordial discs irradiated by a young central star to determine the steady-state solution for the photoevaporative flow. 
The total mass loss rate and the radial mass loss profiles are then obtained from the radiation-hydrodynamical calculations and used as a sink term in a 1D viscous evolution code, in order to predict the surface density evolution of the disc. The three main steps and codes used for each of these calculations are common with previous papers and are only briefly described in the next section, where the references to the more complete descriptions and validation of the methods are also given.

% ----------------------------------------
% ----------------------------------------
% ----------------------------------------
\subsection{Thermal calculations} 
\label{sec:mocassin}

\begin{figure}
\centering
\includegraphics[width=\hsize]{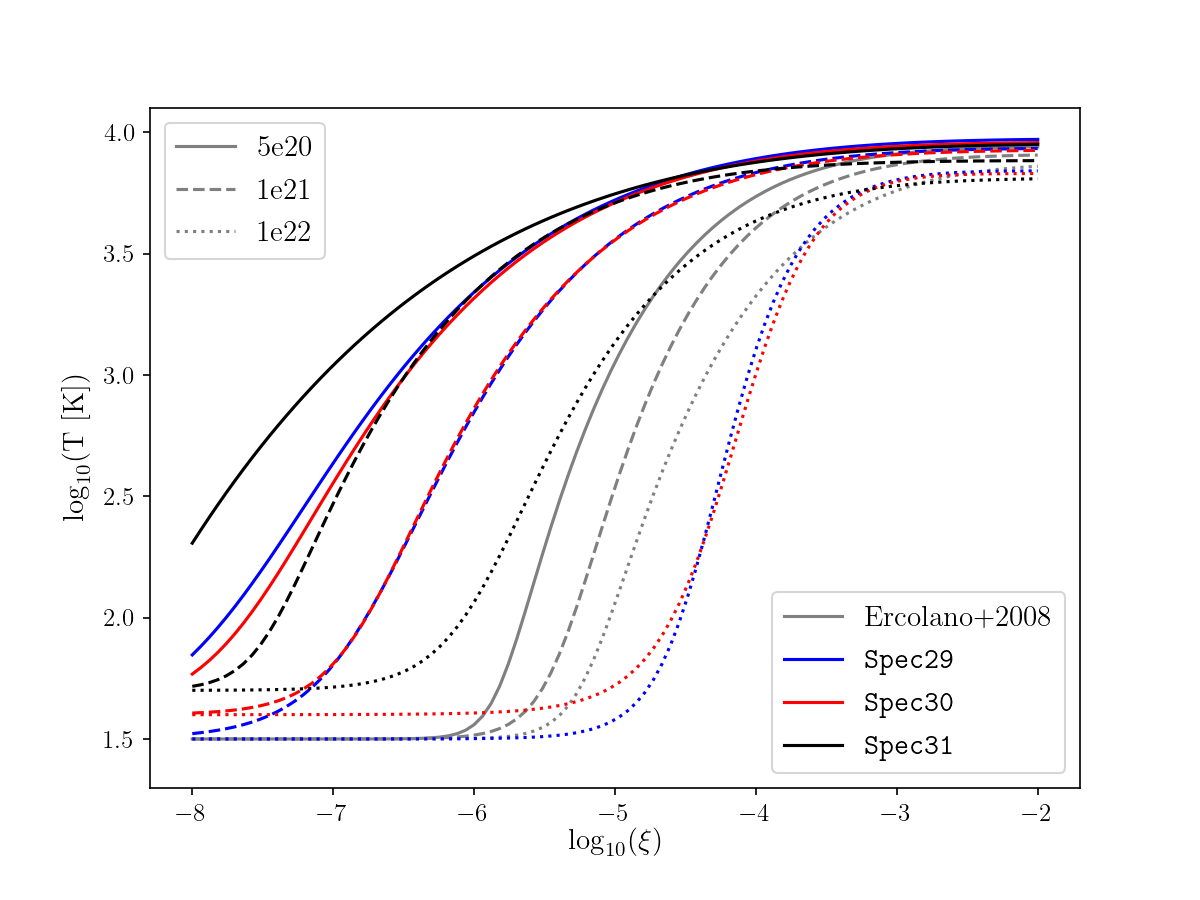}
  \caption{Temperature as a function of ionisation parameter for the X-ray spectrum of \citet{Ercolano+2008b} shown in gray, compared to those derived in this work for $\Lx = 10^{29}\,\ergs$ (blue, \texttt{Spec29}), $\Lx = 10^{30}\,\ergs$ (red, \texttt{Spec30}), and $\Lx = 10^{31}\,\ergs$ (black, \texttt{Spec31}), for three representative column densities, $N=5\times10^{20}, 1\times10^{21}$ and $1\times10^{22}\,\mathrm{cm}^{-2}$.}
     \label{fig:xite}
\end{figure}

\citet{Woelfer_2019} and \citet{Ercolano_2010} showed that for the same ionisation parameter, $\xi=L_X/(n r^2)$ (where $n$ is the volume number density and $r$ is the spherical radius), column density $N$, and irradiation spectrum, the temperature of the gas is sensitive to the local elemental abundances. The irradiation spectrum, however, is also expected to influence the temperature of the gas. We thus follow \citetalias{Picogna2019} and \cite{Woelfer_2019} and use the photoionisation code \mocassin\ \citep{Ercolano+2003, Ercolano+2005, Ercolano+2008a_mocassin} to determine parameterisations of the gas temperature as a function of the ionisation parameter for 40 different gas column densities between $5 \times 10^{20}\,\mathrm{cm}^{-2}$ to $2 \times 10^{22}\,\mathrm{cm}^{-2}$ for our new irradiating spectra. We refer to the previous work mentioned above for details about the \mocassin\ code \citep{Ercolano+2003, Ercolano+2005, Ercolano+2008a_mocassin} and the elemental abundances used \citep{Picogna2019}, and only list the major heating and cooling channels included in the temperature equilibrium calculations in this work. Heating in the wind region and in the disc atmosphere is from photoionisation. Dust photoelectric heating and dust-gas-collisions are included, but unimportant in these regions. Cooling is mainly via collisionally excited and recombination lines from abundant species. Other processes like free-free, Thomson and Compton scattering and two-photon continuum radiation are also included, but play a secondary role. 

In Figure~\ref{fig:xite} we compare the temperature as a function of ionisation parameter for the X-ray spectrum of \citet{Ercolano+2008b, Ercolano+2009} to those derived in
this work for $\Lx = 10^{29}$, $10^{30}$, and $10^{31}\,\ergs$ at three selected column densities of $N=5 \times 10^{20}$, $1\times10^{21}$ and $1\times10^{22}\,\mathrm{cm}^{-2}$. 
 For column densities of $N \sim 5 \times 10^{20}\,\mathrm{cm}^{-2}$ up to the wind launching region of $N \sim 10^{22}\,\mathrm{cm}^{-2}$, the temperature at a given $\xi$ is higher in our new parameterisations, compared to previous work of \citet{Picogna2019}. This is mainly due to the fact that in the new spectra the fraction of energy coming out in the soft X-ray band (0.1 to 1 keV) is over a factor two higher than in the spectrum of \citet{Ercolano+2008b, Ercolano+2009}. Soft X-ray photons are most efficient at heating in this column density range. Local factors, including the specific cooling channels and the ionisation structure affecting may also play a role.

%The new spectra are overall harder than the spectrum of \citet{Ercolano+2008b, Ercolano+2009}, with the result that, for column densities up to the wind launching region $N \sim 10^{22}\,\mathrm{cm}^{-2}$, the temperature at a given $\xi$ is higher in our new parameterisations, as the average energy of the ejected photo-electrons is larger because of the on-average more energetic ionising photons. 

We describe the temperature prescription as a function of $\log{\xi}$, adopting a sigmoidal function as in \citetalias{Picogna2019}:

\begin{equation}
\label{eq:xiTe}
    \log{T_\mathrm{e}} = d + \frac{a-d}{[1+(\log{\xi}/c)^b]^m},
\end{equation}
where the parameters for the different column densities and spectra are given in Tables~\ref{tab:Spec29}, \ref{tab:Spec30}, and \ref{tab:Spec31} for the \texttt{Spec29}, \texttt{Spec30} and \texttt{Spec31} models, respectively.

% ----------------------------------------
% ----------------------------------------
% ----------------------------------------
\subsection{2D Hydrodynamical calculations} 
\label{sec:2dhydro}

We performed radiative-hydrodynamical simulations using a modified version of the \pluto\ code \citep{Mignone_2007, Picogna2019}, to accurately model the stellar irradiation. In the upper layers of the disc (i.e. in the column density range $5\times10^{20}$ to $2\times10^{22}$\,cm$^{-2}$), we adopted the parameterisation presented in the previous section, coming from the radiative transfer calculations in \mocassin. We linearly interpolated the temperatures for intermediate values of column densities and extrapolated the temperature for ionisation parameters outside the simulated range. The regions where these extrapolations were necessary were nevertheless circumscribed. For the bulk of the disc (i.e. column densities larger than $2\times10^{22}$\,cm$^{-2}$), we assumed a perfect coupling between the gas and dust temperature, where the latter was mapped from the models of \citet{D_Alessio_2001}.
For more details on the numerical implementation, see \citetalias{Picogna2019}.

\begin{figure*}
  \includegraphics[width=\textwidth]{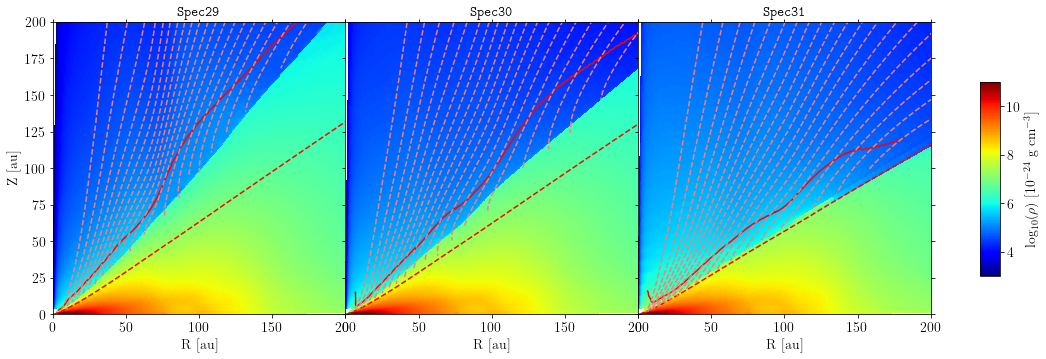}
  \caption{Density distribution at equilibrium for the different spectra. From left to right: \texttt{Spec29}, \texttt{Spec30}, \texttt{Spec31}. The solid red line shows the sonic surface, while in dashed red is plotted the maximum penetration depth of X-rays, and in dashed orange the gas streamlines every 5\,\% of the cumulative mass loss rate.}
  \label{fig:dens2Dspectra}
\end{figure*}

The numerical grid adopted was chosen to maximise the resolution close to the wind launching region. For this region we over-sampled the grid, adopting a static nested grid as summarised in Table~\ref{tab:pluto}.
\begin{table}
\noindent
\centering
\begin{tabular}{cc}
\hline
\hline
\textbf{variable} & \textbf{value} \\
\hline
\textit{disc extent} & \\
%\hline
radial (au) & 0.33--1000 \\
polar (rad) & [0.01--0.5, 0.5--1.0, 1.0--$\pi$/2] \\
\hline
\textit{grid resolution} & \\
%\hline
radial & 412\\
polar & [100, 200, 50]\\
\hline
\end{tabular}
\caption{Input parameters for the 2D hydrodynamical models presented in Section~\ref{sec:2dhydro}.} 
\label{tab:pluto}
\end{table}
The hydrodynamical runs were started from the steady state solutions of \citet{Picogna2019} and then, once the new temperature prescription was applied, evolved for a few hundred orbits at 10 au until a new steady state in the flow was reached.
In Figure~\ref{fig:dens2Dspectra}, we show the gas density distribution for the three different cases, where the gas streamlines are overplotted at 5\,\% interval steps of the cumulative mass loss rate.

The adopted X-ray luminosity has a strong impact on the disc flaring.
For the same gas local properties and column densities, the gas temperature is higher for higher X-ray luminosity. Thus, the stellar irradiation can generate a thermal wind from deeper and denser regions in the disc.
For high X-ray luminosities, we observe a flatter disc with a denser wind region, preventing the radiation from reaching the outermost parts of the disc. For low X-ray luminosities, the radiation is not able to heat the gas sufficiently to unbound it, and therefore the disc shows a larger flaring and a less dense wind. This allows the radiation to remove mass from a larger region of the disc, as one can infer from the resulting surface mass loss profile shown in Figure~\ref{fig:sigmadot}.

% ----------------------------------------
% ----------------------------------------
% ----------------------------------------
\subsection{1D surface density calculations} 
\label{sec:spock_discs}

\begin{table*}
\noindent
\begin{tabular}{ccccc}
\hline
parameter & \citetalias{Picogna2019} & \texttt{Spec29} & \texttt{Spec30} & \texttt{Spec31} \\
\hline
a	 	    & 	-0.5885	    &	-0.9152	    &	0.3034	    &	-1.2845	    \\
b	 	    & 	4.3130	    &	8.5032	    &	-1.5323	    &	9.3601	    \\
c	 	    & 	-12.1214	&	-32.0623	&	1.5766	    &	-27.7371	\\
d	 	    & 	16.3587	    &	62.8336	    &	4.0211	    &	42.9367	    \\
e	 	    & 	-11.4721	&	67.9150	    &	-11.1311	&	-37.3244	\\
f	 	    & 	5.7248	    &	39.2652	    &	10.6550	    &	8.8216	    \\
g	 	    & 	-2.8562	    &	-10.1113	&	-4.5769	    &	-5.3780	    \\
$R_\mathrm{cut}$ (au) & 120. & 206. & 105. & 69. \\
$\Mdotwind~(M_\odot/\mathrm{yr})$ & cf. Eq.~\ref{eq:Picogna_Mdotwind} & $7.0\times10^{-9}$ & $3.9\times10^{-8}$ & $1.3\times10^{-7}$ \\
\hline
\end{tabular}
\caption{Parameterisation for the surface mass loss profile described by Eq.~\ref{eq:Picogna_Sigmadotwind} and the resulting wind-mass loss rates for the different irradiating stellar spectra modelled in our study.} 
\label{tab:params_sigmadotwind}
\end{table*}

The surface density evolution of a disc subject to viscosity and X-ray photoevaporation driven by the host star can be described by the following equation:

\begin{equation} 
    \frac{\partial \Sigma}{\partial t} = \frac{1}{R}\frac{\partial}{\partial R}\left[ 3R^{1/2} \frac{\partial}{\partial R}\left(\nu \Sigma R^{1/2}\right) \right] - \dot{\Sigma}_{\mathrm {w}}(R,t),
\label{eq:sigma_evol}
\end{equation}
where the first term describes the viscous evolution of the circumstellar disc \citep{LyndenBellPringle1974}, and $\Sigmadotwind(R,t)$ corresponds to the surface mass loss profile due to photoevaporation. Here, $\Sigma(R,t)$ describes the gas surface density, $M_\star$ is the stellar mass, $R$ the distance from the central star and $G$ is the gravitational constant.
In order to calculate the global evolution of the disc, we employed the 1D viscous evolution code \spock\ \citep{ER15} and discretised Eq.~\ref{eq:sigma_evol} on a grid of 1000 radial cells equispaced in $R^{1/2}$, which extends from $0.04\,\mathrm{au}$ to $10^4\,\mathrm{au}$. The locally isothermal discs have an aspect ratio of $H/R=0.1$ at $R_1$, which results in flared discs following $H \propto R^{5/4}$, and a midplane temperature scaling as $T_\mathrm{mid}\propto R^{-1/2}$, where $T_\mathrm{mid}\approx[2100\,\mathrm{K}, 4\,\mathrm{K}]$ at the inner and outer boundary, respectively.

Considering a stellar mass of $M_\star=0.7\,M_\odot$ with an initial disc mass of $0.07\,M_\odot$, we set $R_\mathrm{1}=50\,\mathrm{au}$, which defines the exponential cutoff to the surface density and therefore sets the viscous timescale $t_\nu=R_\mathrm{1}^2/(3\nu)$.
The viscosity is defined as $\nu=\alpha c_\mathrm{s}H$, where $c_\mathrm{s}$ is the sound speed of the gas, $H$ the disc scale height and $\alpha$ the dimensionless Shakura-Sunyaev parameter \citep{ShakuraSunyaev1973}. In order to obtain observationally motivated lifetimes ranging from 3--10\,Myr \citep[e.g.][]{Ribas+2014, Ribas+2015} for discs subject to viscosity and X-ray photoevaporation, we chose $\alpha=10^{-3}$.

\begin{figure}
\centering
\includegraphics[width=\hsize]{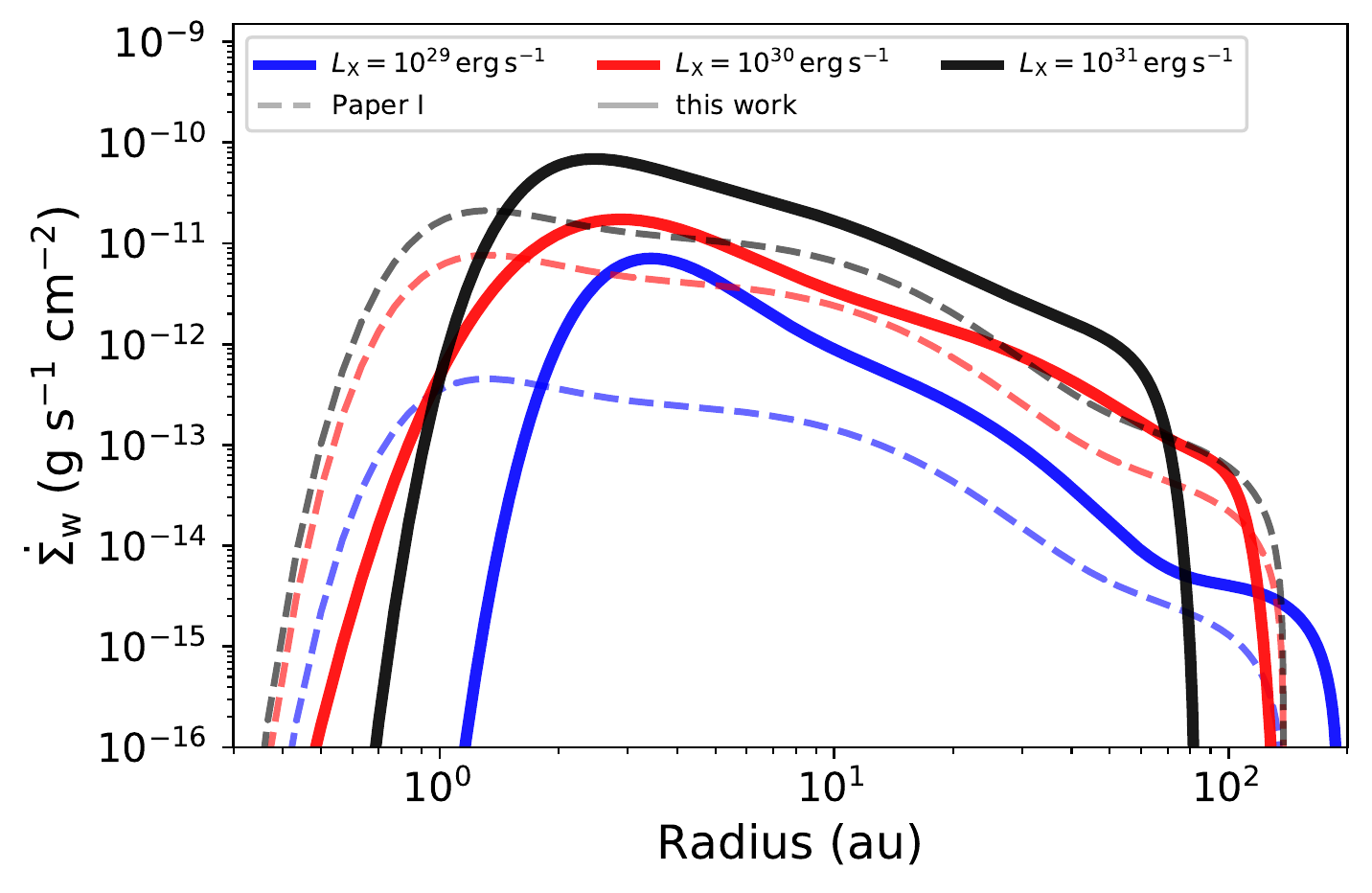}
  \caption{Surface mass loss rate as a function of disc radius obtained for the different irradiating spectra summarised in Table~\ref{tab:params_sigmadotwind}. The solid lines show the results for the \texttt{Spec29} (blue), \texttt{Spec30} (red) and \texttt{Spec31} (black) spectrum presented in this work. The dashed lines show the results for the X-ray input spectrum of \citet{Ercolano2008, Ercolano2009}, which was used in \citetalias{Picogna2019}, evaluated for the same X-ray luminosities.}
     \label{fig:sigmadot}
\end{figure}

Mass loss due to photoevaporation is implemented as a sink-term to the global surface density evolution of the disc, which removes a given amount of gas from each cell at every timestep. The surface mass loss profile from \citetalias{Picogna2019} is described by:

\begin{align}
\label{eq:Picogna_Sigmadotwind}
    \Sigmadotwind (R) =& \ln{10}\,\bigg( \frac{6a\ln{R}^5}{R\ln{10}^6} +\frac{5b\ln{R}^4}{R\ln{10}^5} +\frac{4c\ln{R}^3}{R\ln{10}^4}\\ 
    \nonumber 
    &+\frac{3d\ln{R}^2}{R\ln{10}^3} +\frac{2e\ln{R}}{R\ln{10}^2}
    +\frac{f}{R\ln{10}} \bigg)\, \frac{\Mdotwind(R)}{2\pi R},
\end{align}
where 

\begin{align}
    \Mdotwind(R) =& 10^{a\log{R}^6 + b\log{R}^5 + c\log{R}^4+ d\log{R}^3} \\ \nonumber
    &\times 10^{ e\log{R}^2 + f\log{R} + g} \Mdotwind(\Lx),
\end{align}
and
\begin{equation}
\label{eq:Picogna_Mdotwind}
    \log \Mdotwind(\Lx) = a_\textrm{L}  \exp{\left(\frac{(\ln{(\log{\Lx})} - b_\textrm{L})^2}{c_\textrm{L}}\right)} + d_\textrm{L},
\end{equation}
with $a_\textrm{L} = -2.7326$, $b_\textrm{L} = 3.3307$, $c_\textrm{L} = -2.9868\times10^{-3}$ and $d_\textrm{L} = -7.2580$. 
The prefactors $a$ to $g$ are different for each irradiating spectrum, while the overall shape of the function describing the mass loss is mostly not affected, besides an exponential cut-off term that prevents any mass loss outside the radius $R_\mathrm{cut}$, so that:

\begin{equation}
    \dot{\Sigma}_\mathrm{w, hard}(R) = \Sigmadotwind (R)  \exp\left[-\left(\frac{R}{R_\mathrm{cut}}\right)^{10}\right].
    \label{eq:sigmadotwind_hard}
\end{equation}
The corresponding prefactors and values of $R_\mathrm{cut}$ for each irradiating spectrum are given in Table~\ref{tab:params_sigmadotwind}. Figure~\ref{fig:sigmadot} compares the resulting surface mass loss rate as a function of disc radius, obtained for the different irradiating spectra presented in this work to those of \citetalias{Picogna2019} obtained with the spectra presented by \citet{Ercolano+2008b, Ercolano+2009}.

\subsection{Giant planet migration calculations}
\label{sec:spock_planets}

As a next step, we expanded Eq.~\ref{eq:sigma_evol} in order to include the interaction of an embedded giant planet with its surrounding gaseous disc, which then reads:

\begin{equation} 
    \frac{\partial \Sigma}{\partial t} = \frac{1}{R}\frac{\partial}{\partial R}\left[ 3R^{1/2} \frac{\partial}{\partial R}\left(\nu \Sigma R^{1/2}\right) - \frac{2 \Lambda \Sigma R^{3/2}}{(G M_\star)^{1/2}}\right] - \dot{\Sigma}_{\mathrm {w}}(R,t).
\label{eq:sigma_evol_planet}
\end{equation}
The second term now treats the migration of the planet due to the specific angular momentum transfer from the planet to the disc, which is described by $\Lambda$. 

We adopt the same approach as described in \cite{ER15}, but adjust the employed initial conditions in order to fit the different photoevaporation profiles in our study. Fully-formed planets of random mass between 0.5--$5\,M_\mathrm{J}$ are inserted at 10\,au into the disc and migrate inwards following the impulse approximation \citep[][but see also \citealt{Monsch+2021}]{LinPapaloizou1979, LinPapaloizou1986, Armitage+2002}. While the planets are allowed to accrete mass during their orbital evolution, their formation itself is not treated in our model. 
Therefore the insertion times of the planets were drawn randomly from a uniform distribution between $0.25~\mathrm{Myr}$ and $t_\mathrm{clear}$, where

\begin{equation}
    t_\mathrm{clear} = \frac{t_\mathrm{\nu}}{3} \left( \frac{3 M_\mathrm{d}}{2t_\mathrm{\nu} \dot{M}_\mathrm{w}} \right)^{2/3},
    \label{eq:tclear}
\end{equation}
is the time at which photoevaporation starts to clear the disc \citep{Clarke+2001, Ruden2004}. The lower limit of $0.25\,\mathrm{Myr}$ was chosen arbitrarily in order to provide a reasonable amount of time to form a giant planet following the core accretion paradigm \citep{Pollack+1996}.

We ran in total 1000 simulations, for which $\Lx$ was sampled linearly between $\log (\Lx/\ergs)=28.5$--$31.5$. To account for the different surface mass loss profiles from the different irradiating spectra, this X-ray luminosity range was divided into three bins with $\log (\Lx/\ergs)=[28.5$--$29.5$, $29.5$--$30.5$, $30.5$--$31.5]$, in which the corresponding surface mass loss profile (as given by Eq.~\ref{eq:sigmadotwind_hard} and Table~\ref{tab:params_sigmadotwind}) was used. 
Finally, the linearly sampled values of the X-ray luminosities were weighted following the X-ray luminosity function (XLF) of the COUP sample for pre-main sequence stars of $0.5\,\Msun \leq M_\star \leq 0.9\,\Msun$ \citep[][]{Preibisch+2005}. This step is necessary in order to obtain a realistic distribution of X-ray luminosities in our model, as otherwise the linear sampling of $\log \Lx$ would introduce biased distributions especially towards lower values of the X-ray luminosity.

% ----------------------------------------
% ----------------------------------------
% ----------------------------------------
\section{Results}
\label{sec:results}

In this section we discuss the effects of our updated irradiating spectrum on specific aspects of disc evolution. In particular, we focus on ($a$) the global mass loss rates and the radial mass loss profiles obtained from the steady state 2D hydrodynamic calculations, ($b$) the global one dimensional evolution of the disc surface density and ($c$) the final orbital distribution of giant planets in discs. 

\subsection{Mass loss rates and profiles}
\label{sec:results_mdot}

\begin{figure}
\centering
\includegraphics[width=\hsize]{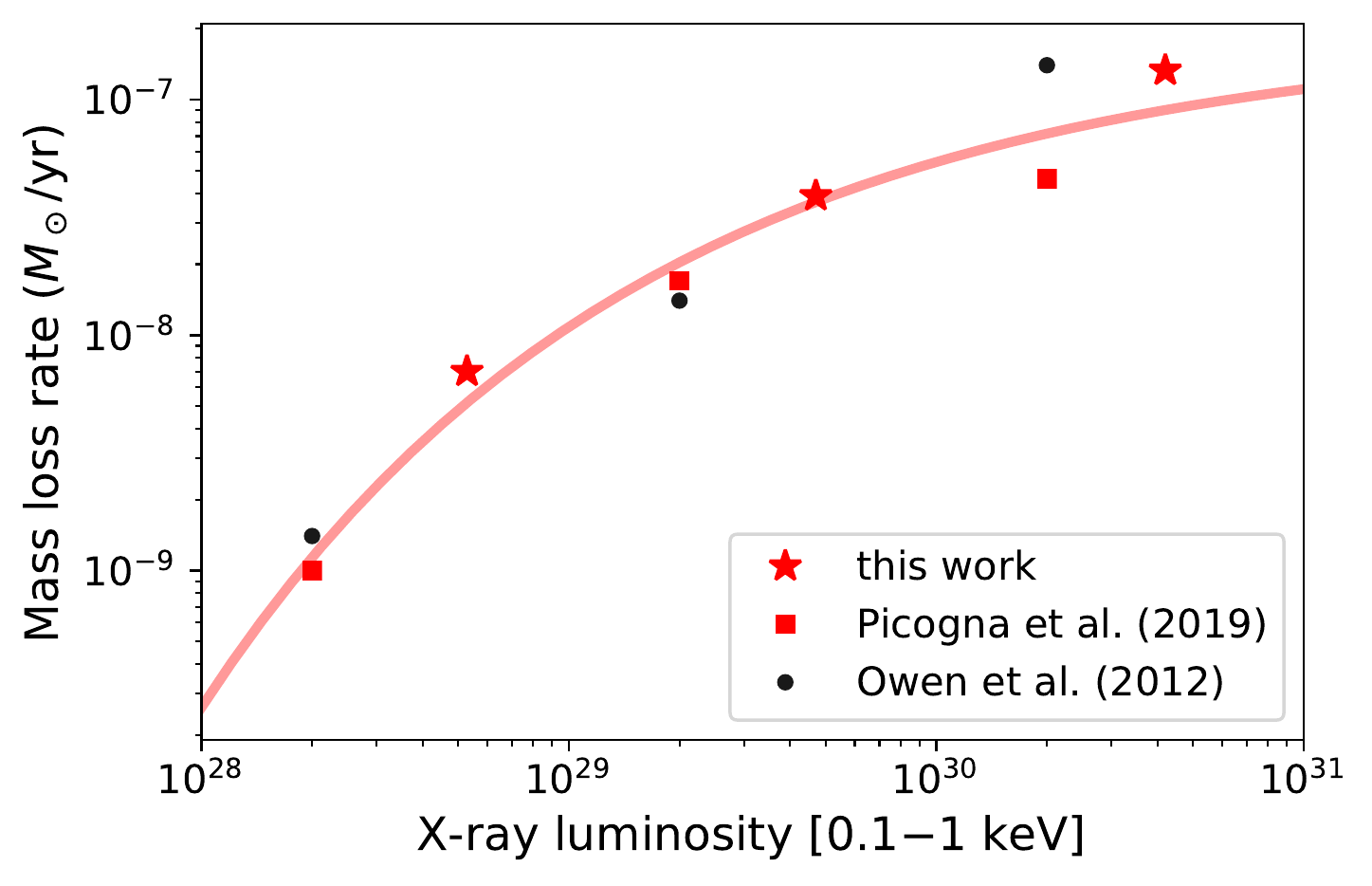}
  \caption{Wind mass loss rate as a function of X-ray luminosity integrated between $100\,\mathrm{eV}$ to $1\,\mathrm{keV}$. Red stars show the values for the cumulative mass loss rates determined for the \texttt{Spec29}, \texttt{Spec30} and \texttt{Spec31} models, respectively, which are summarised in Table~\ref{tab:params_sigmadotwind}. Red squares are the corresponding values for the spectrum used in \citetalias{Picogna2019}. The red line shows an appropriate fit for all red data points, which is given by Eq.~\ref{eq:Mdotwind_soft}. For comparison, the mass loss rates from \citet{Owen_2012}, computed within the same spectral range, are overplotted as well. }
     \label{fig:Mdotwind}
\end{figure}

%Figure~\ref{fig:Mdotwind} shows the resulting values for the cumulative mass loss rate $\Mdotwind(\Lx)$ both from this study and \citetalias{Picogna2019} as a function of the integrated X-ray luminosity within the $0.1$--$1.\,\mathrm{keV}$ spectral band. Both models are in well agreement and therefore we obtained a simple fit by using the relation derived \citetalias{Picogna2019}, which is given in Eq.~\ref{eq:Picogna_Mdotwind} and for which we find $a=-1.947\times10^{17}$, $b=-1.572\times10^{-4}$, $c=-2.866\times10^{-1}$ and $d=-6.694$. The values of $\Mdotwind (\Lx)$ given by \citet{Owen_2012}, who obtained a different dependence of their model on the temperature, are over-plotted for comparison
The global mass loss rates obtained in this work are quantitatively compared in Table~\ref{tab:disc_properties} to the ones obtained by \citetalias{Picogna2019} for the same disc configuration, but the spectrum of \citet{Ercolano+2008b, Ercolano+2009}. 
The differences reported in Table~\ref{tab:disc_properties} are driven by the differences in the integrated luminosity in the `soft' X-ray band (0.1--1\,keV) of the spectrum, which are also reported in the same table. 
Figure~\ref{fig:Mdotwind} shows the resulting values for the cumulative mass loss rate $\Mdotwind(L_\mathrm{X, soft})$ both from this study and \citetalias{Picogna2019} as a function of the integrated X-ray luminosity within the 0.1--1\,keV spectral band. The relation for $\Mdotwind (\Lx)$ given by \citet{Owen_2012}, integrated within the same spectral band, is overplotted for comparison.

When plotted against the luminosity in the `soft' X-ray region, both sets of simulations agree to within factors of a few (see Figure~\ref{fig:Mdotwind}). 
We thus find that the mass loss rates integrated over the whole disc are not very sensitive to reasonable changes in the input spectrum, i.e. within the ranges of the observed X-ray output of the central stars.

We provide an updated $\Mdotwind$ parameterisation in terms of $L_\mathrm{x,soft}\,[0.1$--$1\,\mathrm{keV}$], which can be used in population synthesis models to better describe the wind mass loss rates for different stellar irradiating spectra. 
This is shown as the red line in Figure~\ref{fig:Mdotwind}, which uses the same fitting function as given by Eq.~\ref{eq:Picogna_Mdotwind}:

\begin{equation}
\label{eq:Mdotwind_soft}
    \log \Mdotwind(L_\mathrm{X, soft}) = a_\textrm{L}  \exp{\left(\frac{(\ln{(\log{L_\mathrm{X, soft}})} - b_\textrm{L})^2}{c_\textrm{L}}\right)} + d_\textrm{L},
\end{equation}
however now with changed prefactors of $a=-1.947\times10^{17}$, $b=-1.572\times10^{-4}$, $c=-2.866\times10^{-1}$ and $d=-6.694$. 

These results suggest that reasonable estimates of the total wind mass loss rates for different input X-ray spectra can be obtained by computing the integrated input flux in the 0.1--1\,keV spectral band and applying the simple analytical formula above. 
Our models therefore confirm the suggestion of \citet{Ercolano+2009} that the soft X-ray band is most efficient at driving the bulk of the photoevaporative wind. 

\begin{table}
\noindent
\begin{tabular}{l|ccc}
\hline
$\log(\Lx/\ergs)$  & 29  & 30 & 31 \\  
\hline
\hline
$\Mdotwind$ $[10^{-8}\,M_\odot\,\mathrm{yr}^{-1}]$ &  &  &  \\
%\hline
\citetalias{Picogna2019}   &  0.10 & 1.7 & 4.6 \\
This Work & 0.70 & 3.9 &  13.3  \\
\hline
%$\log(\Lx)\,[\ergs]$  & 29  & 30 & 31 \\  
%\hline
disc lifetime [Myr]   & & & \\
%\hline
\citetalias{Picogna2019} & 19.4 & 2.7 & 1.2 \\
This Work & 13.1 & 1.6 & 0.7 \\
\hline
Time of gap opening [\% of $t_\mathrm{disc}$] & & & \\
%\hline
\citetalias{Picogna2019} & 79 & 75 & 72 \\
This Work & 76 & 69 & 43 \\
\hline
 $L_\mathrm{x,soft}$ $[10^{29}\,\ergs]$ & & & \\
%\hline
\citetalias{Picogna2019} & 0.2 & 2.0 & 20. \\
This Work & 0.53 & 4.7 & 42. \\
\hline
\end{tabular}
\caption{Total mass loss rates in the wind, global disc lifetime and luminosity of the input spectrum between 0.1--1\,keV. The global disc lifetime was obtained assuming a disc of initial mass of $0.07\,\Msun$, a disc scaling radius of $R_1 = 50\,\au$ and $\alpha=10^{-3}$. }
\label{tab:disc_properties}
\end{table}

%In Figure~\ref{fig:sigmadot} we compare the surface mass loss rate as a function of disc radius, obtained for the different irradiating spectra presented in this work (solid lines) to those of \citetalias{Picogna2019} obtained with the spectra presented by \citet{Ercolano+2008b, Ercolano+2009}. 
It follows from Figure~\ref{fig:sigmadot} that the radial extent of the surface mass loss profile decreases for larger X-ray luminosities. As discussed in Section~\ref{sec:2dhydro}, this is due to the fact that for high X-ray luminosities the wind in the inner region screens the outer disc region, preventing their irradiation, when diffuse fields are neglected. As a result, the disc scale height at large radii is much smaller for high X-ray luminosities and the mass loss profile is less extended. This can be more quantitatively assessed by comparing the $R_\mathrm{cut}$ parameter listed in Table~\ref{tab:params_sigmadotwind}, which is $69\,\au$ for $\Lx = 10^{31}\,\ergs$, $105\,\au$ for $\Lx = 10^{30}\,\ergs$ and $120\,\au$ for $\Lx = 10^{29}\,\ergs$. 
In comparison, $R_\mathrm{cut}$ stays constant at roughly $120\,\au$ for all three $\Lx$ cases when using the same spectrum as in \citetalias{Picogna2019}. 
This demonstrates that the variation of $R_\mathrm{cut}$ is driven by the variation in spectral hardness rather than the simple variation of $\Lx$. 
 
This effect also explains the flattening of the mass loss rate to X-ray luminosity relation for high values of the X-ray luminosities (Figure~\ref{fig:Mdotwind}), which was already reported in \citetalias{Picogna2019}. This occurs since $\Mdotwind \propto \int \Sigmadotwind (R) R^2 \mathrm{d}R$, thus mass loss in the outer parts of the disc carries a larger contribution to the total mass loss rate.  

% ----------------------------------------
% ----------------------------------------
% ----------------------------------------
\subsection{Evolution of the disc surface density}
\label{sec:results_discs}

\begin{figure*}
\centering
\includegraphics[width=\hsize]{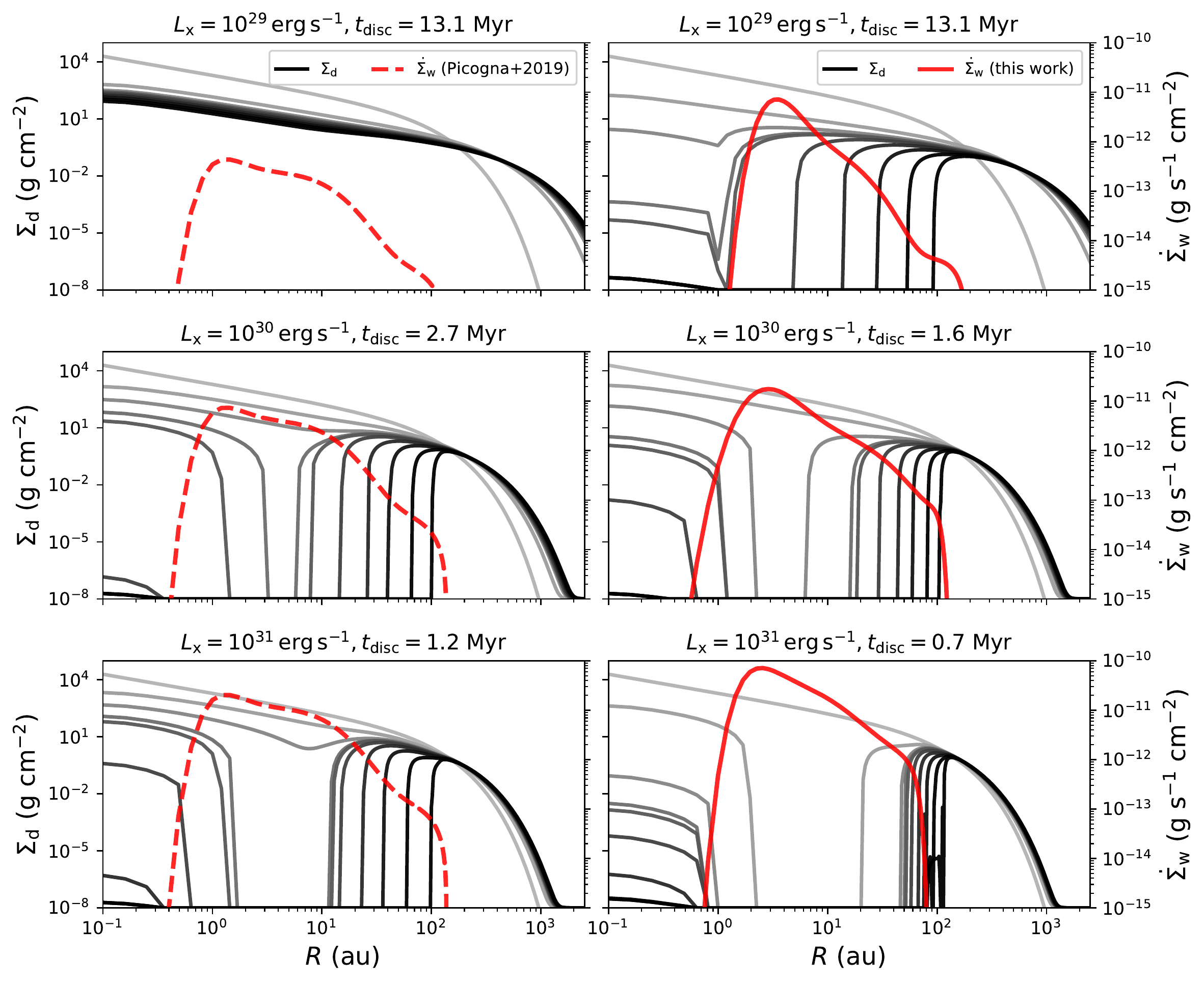}
  \caption{Comparison of the surface density evolution of a disc with initial mass of $M_\mathrm{d}=0.07\,\Msun$, $R_1=50\,\au$, $\alpha=10^{-3}$ using the different input spectra for the X-ray luminosity bins for the calculation of the surface mass loss profile (see Eq.~\ref{eq:Picogna_Sigmadotwind}). The temporal evolution of the disc surface density ($\Sigma_\mathrm{d}$) was plotted at [0, 50, 70, 75, 76, 80, 85, 90, 95, 99]\,\% of the corresponding, total disc lifetime, starting from the lightest grey line and with darker grey lines indicating increasing age.}
     \label{fig:disc_evolution}
\end{figure*}

In this section we explore the global evolution of the surface density of discs subjected to irradiation from the different spectra shown in Figure~\ref{fig:Xray}. 

In Figure~\ref{fig:disc_evolution} the surface density as a function of disc radius is plotted at several time steps of the total disc lifetime, $t_\mathrm{disc}$, for the three spectra used in this work (right columns) and compared to previous calculations by \citetalias{Picogna2019} (left columns).
The red lines show the corresponding mass loss profile and the black lines show the surface density of the disc between [0, 50, 70, 75, 76, 80, 85, 90, 95, 99]\,\% of the corresponding total disc lifetime. This is defined as the time in which the maximum surface density within the disc becomes less than $0.1\,\gcm$ or once the disc reaches a maximum age of $10\,\mathrm{Myr}$, as is the case in the upper left panel, at which the simulation was forced to stop before the disc could be fully dispersed.

When compared to \citetalias{Picogna2019}, the surface density evolution of the disc does not appear to be significantly sensitive to the choice of an observationally-motivated input spectrum. The gap opens roughly at the same location (i.e. between 2--3 au)  and rapidly widens, as the inner disc drains viscously and the outer disc is dispersed away by the photoevaporative flow, powered by direct irradiation. 
While the broad-brush picture is roughly unchanged, there are differences in the global disc lifetimes and in the time at which the gap opens, as can be seen from the values reported in Table~\ref{tab:disc_properties}. These differences can be explained by the fact that we are comparing models that have the same integrated X-ray luminosity, but a different spectral hardness, thus resulting in different luminosities in the soft X-ray band ($L_\mathrm{x,soft}$) in Table~\ref{tab:disc_properties}.

% ----------------------------------------
% ----------------------------------------
% ----------------------------------------
\subsection{Population synthesis: transition disc statistics and giant planet distributions}
\label{sec:results_spock}

Photoevaporation prescriptions play an important role in population synthesis models of protoplanetary discs \citep[e.g.][]{Owen_2011, ErcolanoWeber+2018, Picogna2019} as well as planet formation \citep{Manara_2019} and migration \citep[][]{ER15, Jennings+2018, Monsch+2021}. 
In what follows, we explore whether one can expect appreciable differences in population syntheses of both transition discs or planet migration, by comparing the photoevaporation prescriptions from this work to those of \citetalias{Picogna2019}, using very simple set-ups.

% ----------------------------------------
% ----------------------------------------
% ----------------------------------------
\subsubsection{Transition disc statistics} 
\label{sec:results_TD_stat}

\begin{figure}
\centering
\includegraphics[width=\hsize]{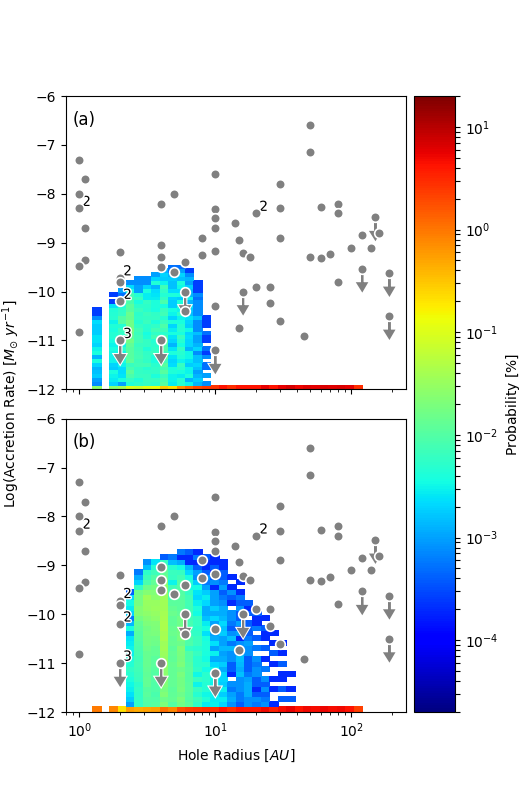}
  \caption{Transition disc demographics of our synthetic populations. The grey circles show observed transition discs from the collection of \citet{ErcolanoPascucci2017}. Numbers next to the circles indicate the total number of sources at that point, arrows indicate that the accretion rate is an upper limit. The coloured areas show the probability of finding a transition disc with the corresponding accretion rate and hole radius, calculated from populations of our X-ray driven photoevaporating disc models. Discs with an accretion rate lower than $10^{-12}$ $M_\odot$ yr$^{-1}$ are shown at the bottom. Panel (a) shows the results for the disc population adopting the old spectra from \citetalias{Picogna2019}, panel (b) shows the results from the new spectra presented in this work.}
     \label{fig:popsynth}
\end{figure}

\begin{figure}
\centering
\includegraphics[width=\hsize]{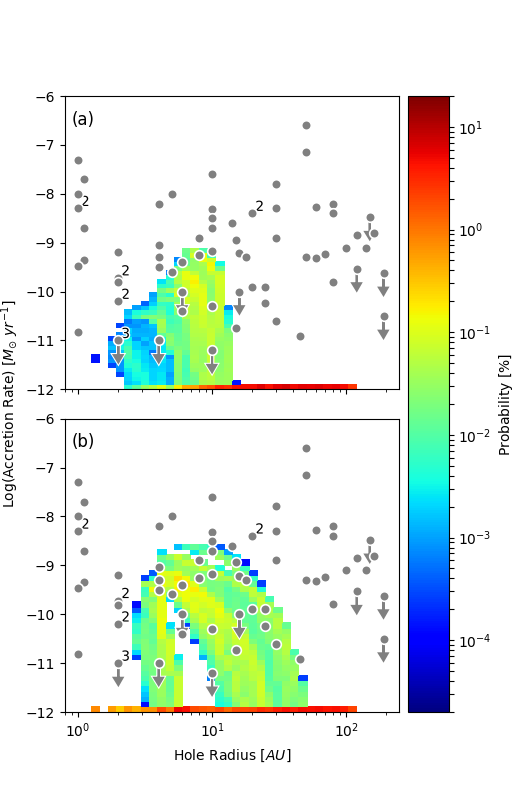}
  \caption{Same as Figure~\ref{fig:popsynth} but with the initial value of $R_1$ randomly sampled as described in \citet{ErcolanoWeber+2018} and the $\alpha$ value chosen in order to retain a disc lifetime consistent with the observed disc population.} 
     \label{fig:popsynth_variable}
\end{figure}

Transition discs, i.e. discs which present an inner cavity depleted of dust, can be formed via several pathways. Giant planet formation and inside-out dispersal from photoevaporative winds are two of the leading theories to explain observations. 
The observed population of transition discs span a large range of central cavity sizes and accretion rates onto the central star. While photoevaporative models are more suited to explain lower rate accretors with small gap sizes, the opposite regime is often associated with the presence of an embedded planet. 
A combination of photoevaporation and supersonic accretion powered by magnetic fields has also been proposed to explain objects with large cavities and large accretion rates, that are difficult to explain otherwise \citep{WangGoodman2017a}. Alternatively, vigorous photoevaporative winds acting in carbon-depleted discs could also help explaining this type of objects \citep{Woelfer_2019}.

In order to understand what fractions of the observed population can be explained by photoevaporation, and what is the influence of the improved spectra presented in this paper, we ran a disc population synthesis (see Figure~\ref{fig:popsynth}), where we compared our new results (panel b) with those coming from the spectra used in \citetalias{Picogna2019} (panel a).
The colour-coding shows the probability of finding a given object at the corresponding accretion rate and hole radius. Due to high amount of objects with low accretion rates and large hole radii, the probability of all remaining objects is considerably low. This can be considered as an artefact of our numerical model and will be discussed in more detail later.
The new spectra are able to cover larger fractions of the observed transition disc population (24.4\,\% vs. 14.6\,\%), by extending the covered part of the plot both towards higher accretion rates ($>10^{-9}$ $M_\odot$/yr), and larger cavity radii ($\sim 40$ au).  

Figure~\ref{fig:popsynth} was obtained adopting a fixed value of the $\alpha$-viscosity for all discs. Choosing a randomly sampled disc scaling radius and adapting the value of $\alpha$ in order to match the observed disc lifetimes \citep[see][for details]{ErcolanoWeber+2018}, we obtain the populations plotted in Figure~\ref{fig:popsynth_variable}, which show an increased probability of observing accreting transition discs. 

Both approaches, however, still fall short of explaining the population of strongly accreting discs with large cavities and further strongly over-predict non-accreting transition discs (so-called `relic discs'). We note, however, that our models only deal with the evolution of the gas phase and it is thus unclear whether the relic discs that are overproduced in our models would actually be observable in the mm-continuum. Indeed, radial drift may have already removed most large grains from the outer regions of these discs by the time the cavity is opened, thus rendering these discs de-facto invisible \citep{ErcolanoJennings2017}. Population syntheses that self-consistently consider dust evolution within the discs are needed to explore this avenue. While these are beyond the scope of this paper, they are the subject of future work. 

 Finally, it is becoming increasingly clear that magneto-hydrodynamic (MHD) winds may also play an important part in the evolution of the surface density of discs, particularly at earlier times \citep[e.g.][]{Kunitomo2020}. Our models do not account for this potentially important component which may have an effect on the resulting population synthesis of transition discs. 

% ----------------------------------------
% ----------------------------------------
% ----------------------------------------
\subsubsection{Giant planet distributions} 
\label{sec:results_planet_popsynth}

\begin{figure}
\centering
\includegraphics[width=\hsize]{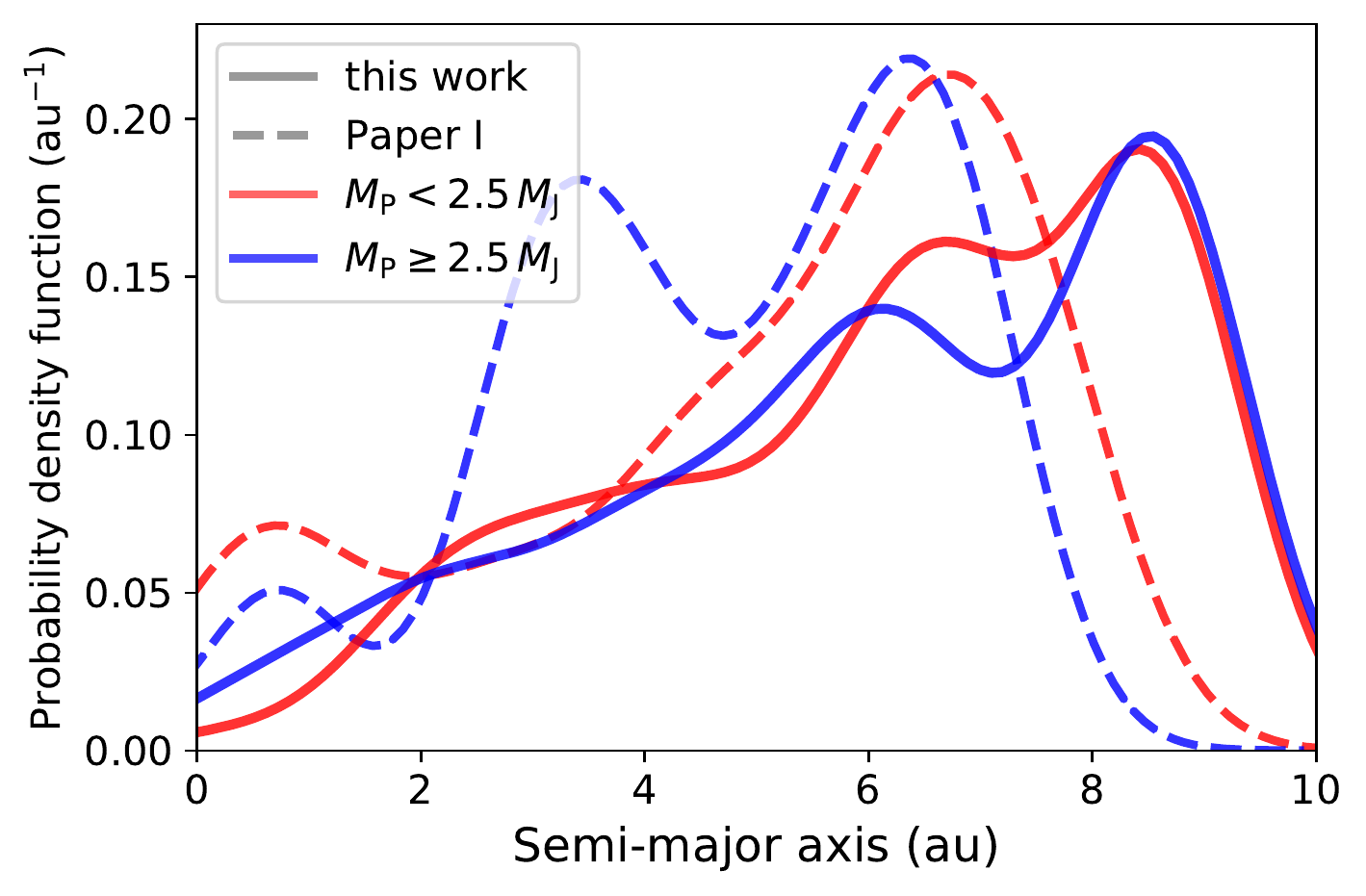}
  \caption{Kernel density estimate for the semi-major axis distribution resulting from the population synthesis models described in Sect.~\ref{sec:results_planet_popsynth}. The models assumed discs with $M_\mathrm{d}=0.07\,M_\odot$, $R_1=50\,\mathrm{au}$ and $\alpha=10^{-3}$. The red lines show the low-mass giant planet population ($M_\mathrm{p} < 2.5\,M_\mathrm{J}$), and the blue lines correspond to the high-mass one ($M_\mathrm{p} \geq 2.5\,M_\mathrm{J}$). The distributions shown as dashed lines follow from the model using the photoevaporation profile as described by \citetalias{Picogna2019}, while the solid lines use the photoevaporation profiles presented in this study. }
     \label{fig:planet_popsynth}
\end{figure}

Figure~\ref{fig:planet_popsynth} shows the results of a planet population synthesis using the same approach as previously described in \cite{ER15} and \citet{Monsch+2021b}. The figure shows the semi-major axis distribution of the planet population synthesis models, calculated using a Gaussian kernel density estimate. The red lines show the low-mass giant planet population ($M_\mathrm{p} < 2.5\,M_\mathrm{J}$), and the blue lines correspond to the high-mass one ($M_\mathrm{p} \geq 2.5\,M_\mathrm{J}$). The distributions shown as dashed lines were obtained using the photoevaporation profile from \citetalias{Picogna2019}, while the solid lines use the profiles obtained in this study.

There are qualitative similarities between the distributions obtained using the \citetalias{Picogna2019} prescription and this work. Both populations predict an under-density of giant planets in the inner disc regions and a peak outside the gap opening regions. However, the position of the peaks and troughs is considerably different. The reason for this is that gap opening occurs earlier in these models (see Figure~\ref{fig:disc_evolution}), thus allowing less time for the planets to migrate inwards before the surface density of the disc around them is depleted by the wind and they are ultimately parked. 

We note that the populations presented here are not meant as a predictive tool to interpret exoplanet survey results. Unfortunately, the lack of a self-consistent planet formation formalism in our 1D code is a serious limitation to the predictive power of our models \citep[cf.][for a detailed discussion]{Monsch+2021b}. However, this exercise is still useful to illustrate the sensitivities of 1D planet population synthesis approaches to the details of the disc dispersal prescription used.

% ----------------------------------------
% ----------------------------------------
% ----------------------------------------
\section{Conclusions} 
\label{sec:conclusion}

We have presented a new set of input X-ray spectra from young stars that was derived from an emission measure analysis of observed T~Tauri stars in the COUP survey. The spectra were used to produce new prescriptions of X-ray photoevaporation, which may be included into disc and planet population synthesis models. The spectra, which are included in the supplementary online data accompanying this paper, can also be used for studies of the irradiation of very young planetary atmospheres at the time of disc dispersal. 

We show that the total mass loss rates ($\Mdotwind$) are controlled by the luminosities in the soft X-ray band $L_\mathrm{x,soft}$ ($100\,\mathrm{eV}$ to $1\,\mathrm{keV}$), and we provide a new relation allowing to scale $\Mdotwind$ by $L_\mathrm{x,soft}$ (Eq.~\ref{eq:Mdotwind_soft}). 
In agreement with previous work presented in \citetalias{Picogna2019}, we find that the total mass loss rates do not scale linearly with X-ray luminosity. At high X-ray luminosities the disc wind in the inner region becomes dense enough to screen the outer disc regions, thus limiting the growth of the total disc mass loss rate. 
This is reflected in the radial profile of the wind, which peaks closer in and does not extend as far for higher X-ray luminosities and harder spectra.

Furthermore, the new spectra are more luminous in the soft X-ray band, resulting in more vigorous wind mass loss rates for the same total X-ray luminosity, compared to the discs irradiated by the spectra presented in \citetalias{Picogna2019}. This strongly affects the surface density evolution of the discs, resulting in a faster dispersal and opening of a gap in more massive discs. 

The agreement between predicted and observed  transition disc statistics moderately improved using the new spectra. However, the observed population of strongly accreting, large-cavity transition discs can still not be reproduced. 
Furthermore, the new models still over-predict the population of non-accreting transition discs.
We suggest, however, that due to radial drift having depleted the outer disc of mm-sized grains, this non-accreting transition disc population may be difficult to observe, thus somewhat easing the tension between the model predictions and the observations.  

Finally we show that simple planet population synthesis models are very sensitive to the choice of the disc dispersal prescription, which further motivates the development of accurate photoevaporative disc dispersal models. 

\section{Data availability}
The data underlying this article are available in the article and in its online supplementary material.

% ----------------------------------------
% ----------------------------------------
% ----------------------------------------
\appendix

\section{Temperature Prescription}

Tables~\ref{tab:Spec29}, \ref{tab:Spec30} and \ref{tab:Spec31} list the parameters for the temperature prescription given in equation~\ref{eq:xiTe}.

\begin{table}
    \centering
    \begin{tabular}{c|c|c|c|c|c|c|c}
         column density & a & b & c & d & m\\
         \hline
         \hline
        5e20 & 1.5 & -18.150 & -7.856 & 3.973 & 0.278 \\
        1e21 & 1.5 & -21.945 & -6.861 & 3.935 & 0.269 \\
        1.5e21 & 1.5 & -17.592 & -6.503 & 3.919 & 0.357 \\
        2e21 & 1.5 & -23.580 & -6.154 & 3.882 & 0.322 \\
        2.5e21 & 1.5 & -24.918 & -6.018 & 3.920 & 0.268 \\
        3e21 & 1.5 & -26.731 & -5.644 & 3.903 & 0.306 \\
        3.5e21 & 1.5 & -34.038 & -5.420 & 3.893 & 0.271 \\
        4.e21 & 1.5 & -26.362 & -5.149 & 3.882 & 0.430 \\
        4.5e21 & 1.5 & -28.736 & -5.130 & 3.877 & 0.374 \\
        5e21 & 1.5 & -22.981 & -4.975 & 3.872 & 0.497 \\
        5.5e21 & 1.5 & -16.510 & -4.816 & 3.872 & 0.700 \\
        6e21 & 1.5 & -16.000 & -4.735 & 3.872 & 0.691 \\
        6.5e21 & 1.5 & -16.000 & -4.718 & 3.877 & 0.608 \\
        7e21 & 1.5 & -16.000 & -4.623 & 3.865 & 0.674 \\
        7.5e21 & 1.5 & -16.000 & -4.540 & 3.860 & 0.700 \\
        8e21 & 1.5 & -16.637 & -4.462 & 3.859 & 0.700 \\
        8.5e21 & 1.5 & -16.772 & -4.387 & 3.855 & 0.700 \\
        9e21 & 1.5 & -16.246 & -4.354 & 3.853 & 0.700 \\
        9.5e21 & 1.5 & -16.000 & -4.362 & 3.851 & 0.686 \\
        1e22 & 1.5 & -22.027 & -4.446 & 3.841 & 0.477 \\
        1.05e22 & 1.5 & -18.575 & -4.364 & 3.845 & 0.572 \\
        1.1e22 & 1.5 & -17.397 & -4.240 & 3.835 & 0.692 \\
        1.15e22 & 1.5 & -17.616 & -4.219 & 3.823 & 0.700 \\
        1.2e22 & 1.5 & -16.000 & -4.270 & 3.832 & 0.647 \\
        1.25e22 & 1.5 & -17.137 & -4.251 & 3.846 & 0.700 \\
        1.3e22 & 1.5 & -29.908 & -4.730 & 3.624 & 0.417 \\
        1.35e22 & 1.5 & -44.599 & -4.702 & 3.625 & 0.280 \\
        1.4e22 & 1.5 & -45.048 & -4.624 & 3.620 & 0.331 \\
        1.45e22 & 1.5 & -31.371 & -4.606 & 3.617 & 0.455 \\
        1.5e22 & 1.5 & -24.526 & -4.553 & 3.629 & 0.559 \\
        1.55e22 & 1.5 & -37.325 & -4.536 & 3.621 & 0.411 \\
        1.6e22 & 1.5 & -100.000 & -4.558 & 3.619 & 0.155 \\
        1.65e22 & 1.5 & -27.527 & -4.459 & 3.615 & 0.555 \\
        1.7e22 & 1.5 & -100.000 & -4.540 & 3.625 & 0.133 \\
        1.75e22 & 1.5 & -21.203 & -4.374 & 3.634 & 0.700 \\
        1.8e22 & 1.5 & -21.639 & -4.344 & 3.628 & 0.700 \\
        1.85e22 & 1.5 & -100.000 & -4.552 & 3.634 & 0.118 \\
        1.9e22 & 1.5 & -100.000 & -4.565 & 3.628 & 0.111 \\
        1.95e22 & 1.5 & -31.118 & -4.328 & 3.629 & 0.512 \\
        2e22 & 1.5 & -58.579 & -4.472 & 3.639 & 0.210
    \end{tabular}
    \caption{Parameters for the temperature prescription in equation~\ref{eq:xiTe} for the $\log{\Lx}=29$ spectra.}
    \label{tab:Spec29}
\end{table}

\begin{table}
    \centering
    \begin{tabular}{c|c|c|c|c|c|c|c}
         column density & a & b & c & d & m\\
         \hline
         \hline
         5e20 & 1.6 & -23.995 & -7.704 & 3.960 & 0.217 \\
        1e21 & 1.6 & -27.727 & -6.859 & 3.927 & 0.210 \\
        1.5e21 & 1.6 & -20.571 & -6.430 & 3.908 & 0.306 \\
        2e21 & 1.6 & -28.534 & -6.047 & 3.864 & 0.279 \\
        2.5e21 & 1.6 & -80.000 & -5.840 & 3.906 & 0.089 \\
        3e21 & 1.6 & -77.487 & -5.473 & 3.874 & 0.124 \\
        3.5e21 & 1.6 & -46.869 & -5.184 & 3.868 & 0.239 \\
        4.e21 & 1.6 & -74.470 & -5.079 & 3.861 & 0.154 \\
        4.5e21 & 1.6 & -25.499 & -4.868 & 3.859 & 0.468 \\
        5e21 & 1.6 & -24.746 & -4.754 & 3.855 & 0.500 \\
        5.5e21 & 1.6 & -20.763 & -4.630 & 3.845 & 0.654 \\
        6e21 & 1.6 & -20.190 & -4.621 & 3.847 & 0.604 \\
        6.5e21 & 1.6 & -18.177 & -4.501 & 3.836 & 0.750 \\
        7e21 & 1.6 & -17.636 & -4.415 & 3.840 & 0.750 \\
        7.5e21 & 1.6 & -16.721 & -4.413 & 3.843 & 0.750 \\
        8e21 & 1.6 & -16.644 & -4.349 & 3.844 & 0.750 \\
        8.5e21 & 1.6 & -18.656 & -4.303 & 3.829 & 0.750 \\
        9e21 & 1.6 & -16.721 & -4.298 & 3.836 & 0.750 \\
        9.5e21 & 1.6 & -17.576 & -4.261 & 3.830 & 0.750 \\
        1e22 & 1.6 & -16.000 & -4.262 & 3.830 & 0.750 \\
        1.05e22 & 1.6 & -16.000 & -4.273 & 3.820 & 0.750 \\
        1.1e22 & 1.6 & -16.000 & -4.304 & 3.821 & 0.671 \\
        1.15e22 & 1.6 & -16.000 & -4.237 & 3.808 & 0.731 \\
        1.2e22 & 1.6 & -16.000 & -4.227 & 3.810 & 0.750 \\
        1.25e22 & 1.6 & -20.873 & -4.295 & 3.806 & 0.650 \\
        1.3e22 & 1.6 & -80.000 & -4.734 & 3.607 & 0.156 \\
        1.35e22 & 1.6 & -21.333 & -4.597 & 3.616 & 0.635 \\
        1.4e22 & 1.6 & -22.201 & -4.520 & 3.604 & 0.750 \\
        1.45e22 & 1.6 & -20.889 & -4.504 & 3.613 & 0.750 \\
        1.5e22 & 1.6 & -17.737 & -4.460 & 3.632 & 0.750 \\
        1.55e22 & 1.6 & -16.000 & -4.520 & 3.635 & 0.750 \\
        1.6e22 & 1.6 & -16.000 & -4.507 & 3.637 & 0.731 \\
        1.65e22 & 1.6 & -16.000 & -4.849 & 3.653 & 0.473 \\
        1.7e22 & 1.6 & -16.000 & -4.507 & 3.626 & 0.729 \\
        1.75e22 & 1.6 & -16.000 & -4.622 & 3.642 & 0.609 \\
        1.8e22 & 1.6 & -16.000 & -4.592 & 3.635 & 0.649 \\
        1.85e22 & 1.6 & -16.000 & -4.479 & 3.638 & 0.727 \\
        1.9e22 & 1.6 & -16.000 & -4.924 & 3.678 & 0.412 \\
        1.95e22 & 1.6 & -80.000 & -4.608 & 3.620 & 0.147 \\
        2e22 & 1.6 & -22.005 & -4.667 & 3.636 & 0.443
    \end{tabular}
    \caption{Parameters for the temperature prescription in equation~\ref{eq:xiTe} for the $\log{(\Lx)}=30$ spectra.}
    \label{tab:Spec30}
\end{table}

\begin{table}
    \centering
    \begin{tabular}{c|c|c|c|c|c|c|c}
         column density & a & b & c & d & m \\
         \hline
         \hline
         5e20 & 1.7 & -99.660 & -8.590 & 3.954 & 0.044 \\
        1e21 & 1.7 & -44.297 & -7.498 & 3.884 & 0.141 \\
        1.5e21 & 1.7 & -23.265 & -6.982 & 3.857 & 0.329 \\
        2e21 & 1.7 & -18.870 & -6.805 & 3.860 & 0.400 \\
        2.5e21 & 1.7 & -16.173 & -6.734 & 3.879 & 0.400 \\
        3e21 & 1.7 & -14.544 & -6.648 & 3.888 & 0.400 \\
        3.5e21 & 1.7 & -17.330 & -6.676 & 3.883 & 0.314 \\
        4.e21 & 1.7 & -23.577 & -6.605 & 3.898 & 0.215 \\
        4.5e21 & 1.7 & -28.298 & -6.450 & 3.884 & 0.191 \\
        5e21 & 1.7 & -38.879 & -6.317 & 3.865 & 0.151 \\
        5.5e21 & 1.7 & -31.700 & -6.342 & 3.860 & 0.176 \\
        6e21 & 1.7 & -22.163 & -6.218 & 3.843 & 0.281 \\
        6.5e21 & 1.7 & -22.438 & -6.270 & 3.845 & 0.256 \\
        7e21 & 1.7 & -36.034 & -6.212 & 3.838 & 0.162 \\
        7.5e21 & 1.7 & -30.500 & -6.203 & 3.831 & 0.193 \\
        8e21 & 1.7 & -47.757 & -6.138 & 3.814 & 0.132 \\
        8.5e21 & 1.7 & -35.389 & -6.131 & 3.834 & 0.159 \\
        9e21 & 1.7 & -32.633 & -6.093 & 3.828 & 0.177 \\
        9.5e21 & 1.7 & -24.455 & -5.991 & 3.812 & 0.260 \\
        1e22 & 1.7 & -24.433 & -6.035 & 3.811 & 0.247 \\
        1.05e22 & 1.7 & -21.125 & -5.995 & 3.807 & 0.290 \\
        1.1e22 & 1.7 & -30.889 & -6.083 & 3.808 & 0.178 \\
        1.15e22 & 1.7 & -28.621 & -6.019 & 3.809 & 0.201 \\
        1.2e22 & 1.7 & -35.809 & -5.997 & 3.795 & 0.171 \\
        1.25e22 & 1.7 & -24.734 & -5.910 & 3.805 & 0.259 \\
        1.3e22 & 1.7 & -15.639 & -5.866 & 3.767 & 0.500 \\
        1.35e22 & 1.7 & -14.278 & -5.932 & 3.765 & 0.500 \\
        1.4e22 & 1.7 & -14.057 & -5.968 & 3.767 & 0.500 \\
        1.45e22 & 1.7 & -13.756 & -5.996 & 3.771 & 0.500 \\
        1.5e22 & 1.7 & -14.176 & -5.946 & 3.760 & 0.500 \\
        1.55e22 & 1.7 & -13.987 & -5.950 & 3.770 & 0.500 \\
        1.6e22 & 1.7 & -13.770 & -5.951 & 3.770 & 0.500 \\
        1.65e22 & 1.7 & -13.806 & -5.961 & 3.766 & 0.500 \\
        1.7e22 & 1.7 & -13.944 & -5.964 & 3.767 & 0.500 \\
        1.75e22 & 1.7 & -12.973 & -6.025 & 3.787 & 0.500 \\
        1.8e22 & 1.7 & -12.469 & -6.049 & 3.788 & 0.500 \\
        1.85e22 & 1.7 & -13.872 & -5.984 & 3.777 & 0.500 \\
        1.9e22 & 1.7 & -14.485 & -5.938 & 3.779 & 0.500 \\
        1.95e22 & 1.7 & -14.530 & -5.942 & 3.776 & 0.500 \\
        2e22 & 1.7 & -14.535 & -5.910 & 3.783 & 0.500 \\
    \end{tabular}
    \caption{Parameters for the temperature prescription in Eq.~\ref{eq:xiTe} for the $\log{\Lx}=31$ spectra.}
    \label{tab:Spec31}
\end{table}

\bibliographystyle{mn2e}
\bibliography{references}

\section*{Acknowledgements}
We thank the anonymous referee for a detailed and constructive report which helped improve our manuscript. 
We acknowledge the support of the DFG priority program SPP 1992 ``Exploring the Diversity of Extrasolar Planets'' (DFG PR 569/13-1, ER 685/7-1) \& the DFG Research Unit ``Transition discs'' (FOR 2634/1, ER 685/8-2). 
This reasearch was supported by the Excellence Cluster ORIGINS which is funded by the Deutsche Forschungsgemeinschaft (DFG, German Research Foundation) under Germany's Excellence Strategy - EXC-2094 - 390783311. The simulations have been partly carried out on the computing facilities of the Computational Center for Particle and Astrophysics (C2PAP). JJD was funded by NASA contract NAS8-03060 to the \textit{Chandra} X-ray Center and thanks the Director, Pat Slane, and the CXC science team for continuing advice and support.

\end{document}